# Title: Device for detection of activity-dependent changes in neural spheroids at MHz and GHz frequencies

Saeed Omidi[1], Gianluca Fabi[2], Xiaopeng Wang[2], James C. M. Hwang[2*], Yevgeny Berdichevsky[1,3**]

[1]Department of Bioengineering, Lehigh University, Bethlehem, USA
[2]Department of Material Science and Engineering, Cornell University, Ithaca, USA
[3]Department of Electrical and Computer Engineering, Bethlehem, USA

[*]jch263@cornell.edu
[**]yeb211@lehigh.edu

## Abstract

Intracellular processes triggered by neural activity include changes in ionic concentrations, protein release, and synaptic vesicle cycling. These processes play significant roles in neurological disorders. The beneficial effects of brain stimulation may also be mediated through intracellular changes. There is a lack of label-free techniques for monitoring activity-dependent intracellular changes. Electromagnetic (EM) waves at frequencies larger than $1 \times 10^6$ Hz (1 MHz) were previously used to probe intracellular contents of cells, as cell membrane becomes "invisible" at this frequency range. EM waves interact with membranes of intracellular organelles, proteins, and water in the MHz – GHz range. In this work, we developed a device for probing the interaction between active neurons' intracellular contents and EM waves. The device used an array of grounded coplanar waveguides (GCPWs) to deliver EM waves to a three-dimensional (3D) spheroid of rat cortical neurons. Neural activity was evoked using optogenetics, with synchronous detection of propagation of EM waves. Broadband measurements were conducted in the MHz-GHz range to track changes in transmission coefficients. Neuronal activity was found to reversibly alter EM wave transmission. Pharmacological suppression of neuronal activity abolished changes in transmission. Time constants of changes in transmission were in the seconds – tens of seconds range, suggesting the presence of relatively slow, activity-dependent intracellular processes. This study provides the first evidence that EM transmission through neuronal tissue is activity-dependent in MHz – GHz range. Device developed in this work may find future applications in studies of the mechanisms of neurological disorders and the development of new therapies.

## 1. Introduction

Many label-free methods to study the neural activity of the brain in real-time have been developed. These methods include various forms of direct electrical or magnetic recordings of neural activity, as well as proxy methods such as functional magnetic resonance imaging (fMRI) and functional Near-Infrared Spectroscopy (fNIRS) that measure the consequences of neural activity on oxygenation and hemodynamics. However, there is a lack of methods that would allow label-free, nondestructive measurement of the intracellular consequences of neural activity. In this work, we have addressed this gap by developing a novel method to probe intracellular contents of neurons via electromagnetic (EM) waves with MHz – GHz frequencies.

Activity-induced intracellular changes play important roles in a variety of neurological disorders. For example, dysregulation of intracellular $Cl^-$ may lead to seizures and epilepsy (Auer et al., 2020). Seizures can also cause significant increases in intracellular pH (Raimondo et al., 2015). Alzheimer's disease is characterized by the formation of amyloid (Aβ) plaques and tangles composed of misfolded, hyperphosphorylated tau proteins. Recent findings suggest that dysregulated neural activity may be the cause of AD neuropathology, and pathology may in turn cause abnormal function of neural circuits (Grieco et al., 2023). Alpha-synuclein (α-syn) protein plays a critical role in Parkinson's disease. α-syn pathology and neural activity have a bidirectional relationship (Wu et al., 2020), potentially through changes in synaptic vesicle dynamics (Calabresi et al., 2023). Brain stimulation is used to treat symptoms and improve cognition in patients. Mechanisms of beneficial effects produced by stimulation include changes in neurotransmitter release, depletion of synaptic vesicles, and structural synaptic changes (Barbati et al., 2022; Farokhniaee and McIntyre, 2019; Vasu and Kaphzan, 2022). A method to measure these neural activity-induced intracellular changes may find applications in the diagnosis of neurological diseases, monitoring of disease progression, as well as assessment of treatment and neuromodulation strategies.

Electromagnetic waves interact with biological matter by inducing dielectric polarization and displacement, and movement of mobile charges. The nature of these interactions with tissues is frequency dependent: at low frequencies, the polarization of cell membranes predominates, while at high frequencies (> 1 MHz), the cell membrane becomes electrically "invisible" and cell contents, including organelles, vesicles, and cytoplasm, interact with the externally applied electric field. Mechanisms of interaction that play a significant role at high frequencies include polarization of organelle and vesicle membranes, as well as dipolar relaxation of proteins and bound (hydration) water molecules (Foster and Schwan, 1989). Concentrations of ions such as $Na^+$, $K^+$, and $Cl^-$ also play a significant role through modifying water dipolar relaxation (Levy et al., 2012). High-frequency interactions have been used to sense protein conformation changes and dynamics (Ermolina et al., 2015) and to determine the properties of synaptic vesicles in aqueous solutions. Dielectric dispersion of rat brain synaptosomes was in the 1 – 100 MHz range and ascribed to the polarization of the synaptosome membrane (Irimajiri et al., 1975). Dielectric properties of the synaptosome interior represent the properties of the dense suspension of synaptic vesicles and were predicted to depend on the volume fraction of the vesicles in the synaptosome. Altogether, the measurement of high-frequency interactions between electromagnetic waves and neurons may reveal activity-dependent intracellular changes.

To determine the feasibility of this measurement, we developed a device that delivered high-frequency EM waves to neurons. Robust control of neural activity in the measurement sample is critical for reproducible measurement. Three-dimensional (3D) spheroid, or aggregate, cultures contain tens or hundreds of primary cortical neurons in a synaptically-connected network (Dingle et al., 2015). Control of neural activity in these aggregates can be accomplished through optogenetics: activation of a channelrhodopsin expressed in neurons via light pulses (Boyden et al., 2005). To ensure robust control, evoked neural activity can be monitored optically using changes in the fluorescence of a genetically-encoded calcium indicator (Ming et al., 2020). Cortical aggregates can be cultured on adherent substrates (Hasan et al., 2019), enabling integration with microfabricated EM wave delivery devices. Finally, cortical aggregates represent an in vitro model system that is similar to cortical tissue in terms of cellular composition, density, and neural circuitry (Hasan and Berdichevsky, 2021). For the reasons outlined above, surface-adherent cortical spheroids were used in this work.

Coplanar waveguides (CPWs) have been used to deliver EM waves in the GHz range to biological samples. CPWs are patterned on flat surfaces, and propagation of the electric field is affected by the material both below and above the waveguide, to a depth determined by waveguide gaps (Simons, 2004). This enables high-frequency measurements in microscale biological samples placed on top of the waveguides (Mertens et al., 2023). This methodology has been recently used to investigate the cellular response to electrochemotherapy (Tamra et al., 2022), determine electrical properties of a cell nucleus (Du et al., 2022; Ferguson et al., 2023), identify and monitor the development of cancer cells (Lei et al., 2021), study the adhesion, growth, and differentiation of neural cells (Elghajiji et al., 2021; Krukiewicz, 2020), detect rapid cell dynamics like micromotions (Bounik et al., 2022), differentiate live and dead cells (Hwang, 2021), and detect oxidative stress (Ferguson et al., 2021).

In this work, we developed a novel grounded CPW (GCPW) array-based device that integrated optical control of neuronal activity with broadband (MHz – GHz) measurement. The overall goal was to provide the first proof that activity-dependent changes can be detected via EM wave-neuron interaction. Optical stimulation and detection of neural activity were used instead of electrical stimulation and detection to avoid interference with high-frequency electrical measurements. The frequency range was designed to specifically target intracellular changes.

## 2. Methods

### 2.1 Fabrication and test of the measurement device

An 8-channel electrode array comprising eight 50-Ω grounded coplanar waveguides (GCPWs) on a borofloat-33 glass wafer 100 mm in diameter and 500 µm in thickness was fabricated (Figure 1a). The electrodes were made of 15-nm-thick Ti, and 0.2-µm-thick Au in the central region of 100 µm × 100 µm and 0.5 µm outside. The backside of the wafer was coated with 0.1-µm-thick indium tin oxide, which served as a transparent ground plane to allow transmission microscopy. In the central region, the center electrode of the GCPW was 5-µm wide with a 1-µm gap from the ground electrodes. The width and gap were tapered to 2 mm and 240 µm, respectively, at the edge of the wafer. The parallel layout at the center of the device was transitioned to a radial layout at the periphery. In the central region, a 2.5-µm-thick SU8 layer was coated on the electrodes and patterned openings of 100-µm-diameter served as contact zone. The electrode array was positioned below the cellular compartment. The bottom of the glass wafer was supported by a 1.6 mm-thick donut-shaped printed circuit board. The inner and outer diameters of the board were 40 mm and 100 mm, respectively. Figure 1A shows the actual images of the chip: front view (A(i)), backside view (A(ii)), and with a mounted petri dish (A(iii)). Moreover, a phase contrast image of the measurement area including plated neural spheroid on GCPWs, measurement contact zone, and light stimulation zone is shown in Figure 1A(iv). A high-magnification image of the central region of the device containing the sensing portion of the GCPWs is shown in the inset in Figure 1A(iv). Figure 1B shows the cross sections of the chip and a zoomed-in area of waveguides. For the 2-port wideband measurement of each GCPW, sixteen SMA connectors were attached to the edge of the wafer. For multiplexing between the eight GCPWs, the SMA connectors were connected to a Keysight E5092A configurable multiport test set and a Keysight E5080A vector network analyzer (VNA) with a bandwidth of 9 kHz to 9 GHz. Calibration was then carried out by using an electronic calibration module (ECal, Keysight N7552A). This module simplifies calibration process by automatically switching between different standards that the VNA software uses to compute correction factors. This moves the reference planes from the VNA to the SMA connectors, effectively removing the impact of cables and connectors on measurement. A diagram of S-parameters measurement from a device under test is shown in Figure 1C. Figure 1D shows a schematic of the electrical measurement setup, consisting of the chip and the VNA device connected with two cables (2 ports of the device connected to two SMA connectors of one waveguide).

Two-port S parameters were measured from each GCPW and compared with simulation results obtained by the high-frequency software simulator (HFSS) from Ansys. Figure 1E shows the reflection coefficient ($S_{11}$) and transmission coefficient ($S_{21}$) of a typical GCPW with and without the culture medium. Without the medium, the insertion loss is less than 15 dB, and the return loss is greater than 5 dB across the 9-GHz bandwidth. Without the medium, the adjacent-channel crosstalk is less than −20 dB and the difference between different pairs of channels is less than a few decibels. With the medium, the insertion loss increases to 35 dB while the return loss decreases to 3 dB. Assuming the culture medium has a dielectric constant $\varepsilon_R = 79$ and a loss tangent $\tan\delta = 0.5$, the simulated $S_{11}$ and $S_{21}$ generally agree with measured parameters. Figure 1E shows the aforementioned measured and simulated results. Differences in transmission and reflection between GCPWs on the same device were minimal (Fig. 1F), indicating fabrication robustness.

**2.2 Neuronal spheroid preparation and plating on measurement device**

Primary rat cortical neurons were dissociated after dissecting the brain of a post-natal-day 0 or 1 rat pup (Sprague-Dawley, Charles River) and isolating the cortices according to the protocol developed by Brewer et. al (Brewer et al., 1993). All animal use protocols were approved by the Institutional Animal Care and Use Committee (IACUC) at Lehigh University and conducted in accordance with the United States Public Health Service Policy on Humane Care and Use of Laboratory Animals. After cortices' digestion, trituration, and centrifugation, the cell pellet was resuspended in a neural culture medium consisting of 97.45% Neurobasal-A, 0.25% GlutaMAX (100x), 0.3% 30 µg/ml gentamicin, and 2% B27-supplement (Gibco, ThermoFisher). A 9 × 9 array micro-mold (Microtissues Inc.) was used for casting agarose gels to form neural aggregation. These agarose gels were cast one day before cell seeding according to the manufacturer's protocol. A $6.5 \times 10^5$ cells in 190 µl cell suspension was seeded in each agarose gel and allowed to settle down for 15 minutes in 37 ºC, 5% $CO_2$ incubator, then submerged under neural plating medium (10% fetal bovine serum (FBS, Gibco) in Neurobasal-A with 0.5 mM GlutaMAX and 30 µg/ml gentamicin) and incubated for 45 minutes. The plating medium was then replaced with the culture medium and changed every 3 days. Neurons aggregated and formed neuronal spheroids after 24 to 48 hours. Figure S1 shows a phase contrast image of the formed spheroids in the agarose gel. To prepare the measurement device for plating, a 35 mm cell culture petri dish with a 20 mm opening in the middle (Cellvis) was glued on the chip using silicone glue (Silbione). Then, the surface of the measurement area of the device was coated with Poly-D-Lysine (PDL, Sigma Aldrich) overnight to promote neuronal adhesion and washed before plating spheroids. On the day of plating, spheroids were released from the agarose gel and one spheroid was carefully transferred with a 1000 μl pipette tip and to the test chip. Spheroid size was 280.4 +/- 21.47 μm. Figure 2A shows a layer of confocal image for a representative spheroid at the plane where the measurement happens, confirming the presence of neuronal soma and neurites. Figure 2B shows a schematic of 3D neuronal spheroid plated on the waveguides and the cross-section view showing electric field penetration of the cell aggregate. Neural spheroids plated on test chips were characterized by similar power, frequency, and amplitude of spontaneous activity as spheroids plated in regular culture dishes (Figure S2A). Viability of spheroids plated on test chips was also not significantly different than viability of spheroids plated on regular culture dishes (Figure S2B). These findings demonstrate biocompatibility of the measurement chip.

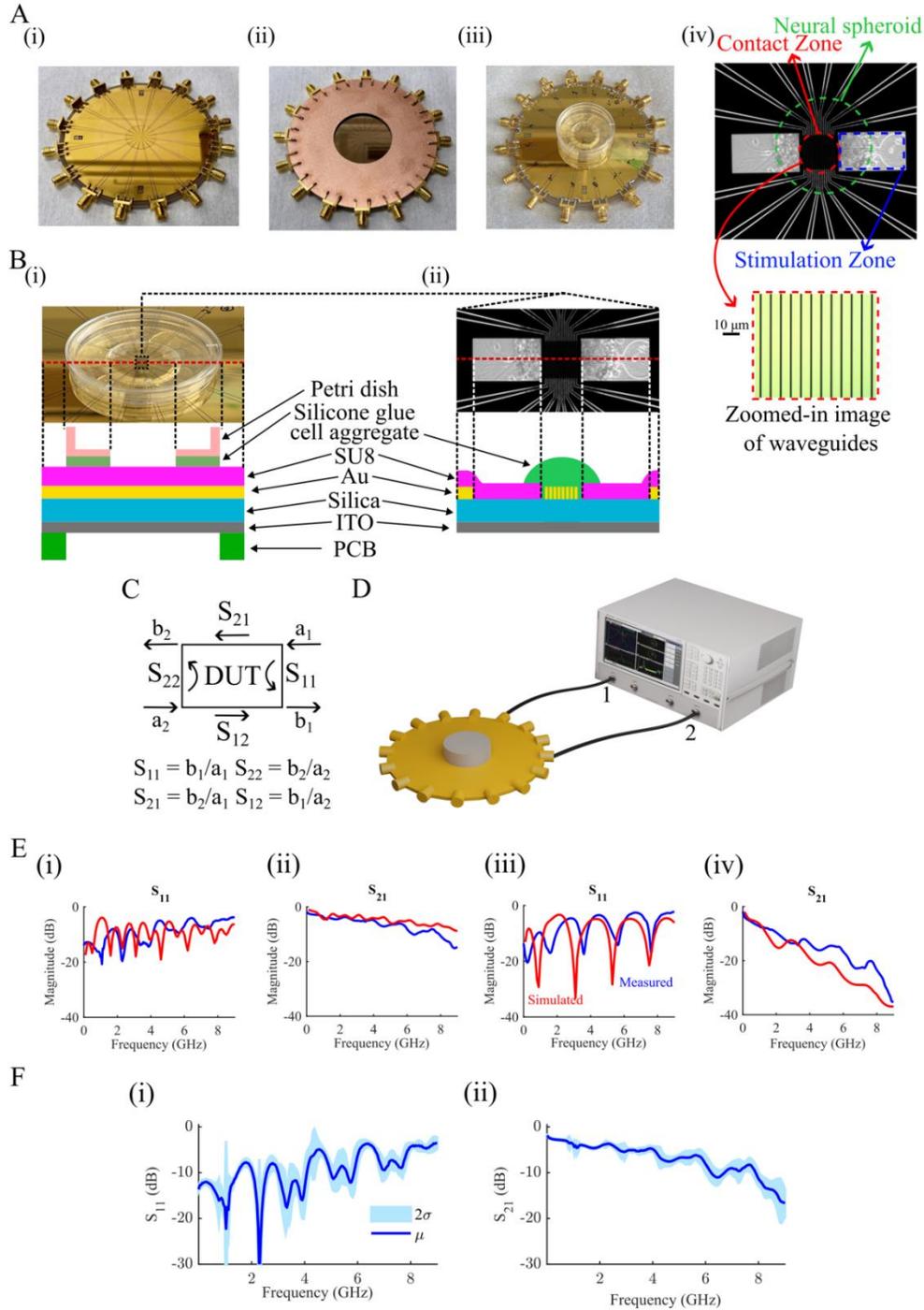

**Figure 1**. Chip fabrication and characteristics. **A.** Photographs of the fabricated device from (i): front view, (ii): backside view, and (iii): with mounted petri dish. A (iv) a phase contrast image of the measurement area showing a plated neural spheroid on top of waveguides, stimulation zone, and measurement contact zones with a zoomed-in image of waveguides. **B.** shows the cross-section of the chip at two scales, (i): the fabricated layers under the peri dish, and (ii): a zoomed-in area corresponding to a square outlined by a dashed line in (i) showing the cross section containing GCPWs and the surrounding area. **C.** A diagram of S-parameters calculation based on the reflected and transmitted power ratios. **D**. Schematic of electrical measurement experimental setup including a VNA device connected to the chip. This 2-port measurement setup corresponds to the diagram in C. **E.** Measured (blue) and simulated (red) reflection ($S_{11}$) and transmission ($S_{21}$) parameters without (i, ii) and with (iii, iv) culture medium. **F**. Average (μ) +/- standard deviation (σ) S parameters of all waveguides on a chip.

## 2.3 Viral infection, calcium recording, and optogenetic stimulation

On day-in-vitro (DIV) 1, the culture medium was replaced with a culture medium containing adeno-associated virus (AAV) particles that express via Synapsin promoter the genetically encoded $Ca^{2+}$ indicator, jRGECO1a (Addgene # 100854), and light-sensitive transmembrane Chanelphodopsin-2 (Addgene # 26973) for optogenetic stimulation. Figure 2C shows a schematic of all-optical control of neural activity: activity was stimulated by shining blue light on neurons expressing optically activated channels (channelrhodopsin-2) and monitored by detecting fluorescence of the calcium indicator jRGECO1a (580/610 nm excitation/emission). A CCD camera (Thorlabs) was used for fluorescent neuronal activity recording with 20 fps frame rate. After approximately 7 days of viral infection, jRGECO1a fluorescent changes indicating the presence of spontaneous neuronal activity were observed. Stimulation was achieved by triggering a light source to deliver blue light pulses through a pattern illuminator Polygon400 (Mightex) to illuminate the region of interest (ROI) in a neuronal spheroid. The stimulation light wavelength was 480 nm, and the illumination power was 10 mW/mm$^2$. The experiment was performed by placing the device with spheroid on the stage of a dual deck inverted fluorescence microscope (IX73, Olympus), and a 10x objective was used for imaging. Figure 2D(i) shows the optical recording/stimulation experimental setup and a zoomed-in image (2D(ii)) shows the spheroid formation, transfer to the device, and stimulation and recording on the device. The device was placed in a mini-incubator mounted on the microscope for the duration of the experiment, to maintain the 5% $CO_2$ and 37 °C environment with a gas cap connected to a humidified mixed blood gas (Airgas) and a temperature controller (Warner Instruments). Modulation stimulation pulses and the recorded video frames during modulation and rest are shown in Figure 2E. The following parameters were used for the optogenetic stimulation waveform: $t_{pt}$ (pulse train duration) is 60 seconds, $t_{rest}$ (rest state gap between modulation events) is 120 seconds, $t_{on}$ is 100 ms, $t_{off}$ is 400 ms, and the number of pulses in the train was 120.

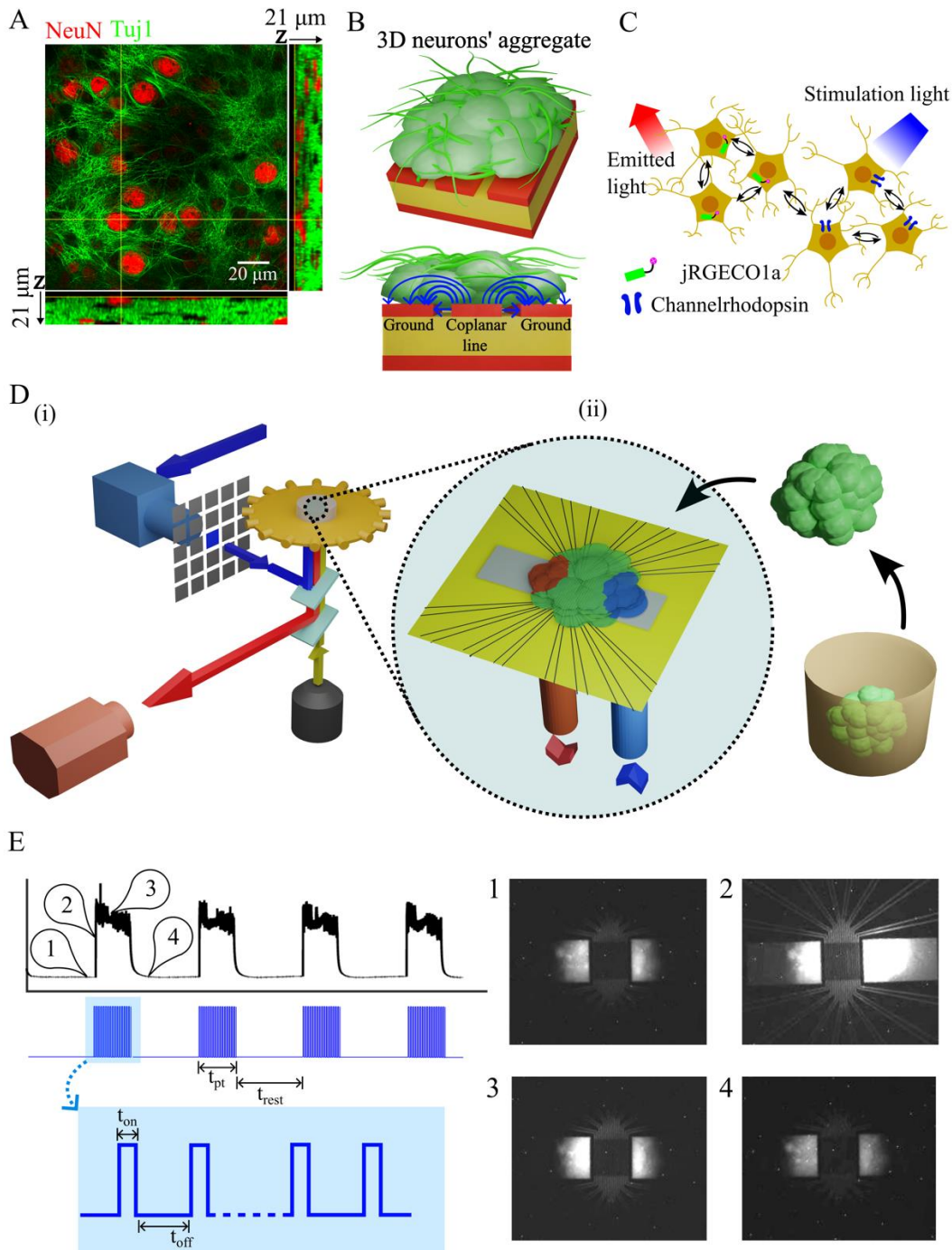

**Figure 2**. Neural spheroid positioning on the device and optical stimulation and detection of neural activity. **A**. A confocal image of a neural spheroid in the plane of electromagnetic measurement. NeuN[+] cells (neurons) are shown in red and Tuj1[+] processes (neurites) are shown in green. **B**. Schematic of the 3D neural spheroid on waveguides. Distribution of the electromagnetic field (blue lines) from GCPWs is shown relative to the cells in the plated spheroid. **C**. Schematic of optogenetic stimulation and $Ca^{2+}$ recording of activity in a network of neurons (black arrows represent synaptic connections). **D**. (i) Optical experimental setup including the blue light pulses passing through the patterned illuminator polygon to an ROI of culture, and emitted light (red) recorded by the camera. (ii) Transfer of a spheroid from the agarose well to the measurement device, and zoomed-in schematic of the measurement area, the plated spheroid, stimulation, and emission light. **E**. A representative sample of neural activity ($\Delta F/F$) evoked by the stimulation protocol (blue pulses). Parameters of stimulation light waveform were: $t_{pt}$ = 60 seconds, $t_{rest}$ = 120 seconds, $t_{on}$ = 100 ms, and $t_{off}$ = 400 ms. On the right, 4 frames of recorded video show 1) rest state pre-stimulation, 2) stimulation light pulse illuminating the culture, 3) neural activity during stimulation, 4) rest state post-stimulation.

## 2.4 Neuronal activity trace calculation and removing light artifacts

Raw mean grey values were extracted from recorded videos using FIJI ImageJ software, and an asymmetric least square smoothing method (Eilers and Boelens, 2005) was used to find the $F_0$ for baseline detection. This method assumes that there is a smooth baseline with a superimposed signal that carries information. The method uses a smoother and an asymmetric weighting of deviations from the smooth trend to effectively estimate the baseline $F_0$. Eq. 1 shows the $\Delta F/F$ calculation. For better visualization, stimulation light pulse artifacts were removed from the $\Delta F/F$ trace.

$$\frac{\Delta F}{F} = \frac{F - F_0}{F_0} \qquad (\text{Eq. 1})$$

## 2.5 S-parameters measurement

Scattering parameters are a matrix of the power ratio reflected and transmitted from the system under test. VNA was used to measure the scattering parameters with the spheroid plated on top of GCPWs by sweeping through the 9 KHz to 9 GHz frequency spectrum. VNA was triggered every 0.5 seconds via TCP/IP connection twice, the first trigger to sweep and the second one to save the measured data once sweeping finished. The collection of data files was read in MATLAB and $S_{11}$ (reflection) and $S_{21}$ (transmission) were extracted as complex numbers. The magnitude of $S_{11}$ and $S_{21}$ were calculated (Eq. 2) and saved as a matrix with time and frequency domains and $|\Delta S_{nm}^{raw}|$ values. A baseline detection method similar to that used for $\Delta F/F$ calculation was used here on every vector of $|\Delta S_{nm}^{raw}|$ for each frequency point. Transmission data ($|\Delta S_{21}^{raw}|$) were smoothed out in the frequency domain using a 5-point moving average algorithm. Then, data was chopped into rest and stimulation periods. The last 20 seconds of stimulation was named 'ON' region, and last 20 seconds of rest before stimulation was called 'OFF' region. $|\Delta S_{21}|$ calculated as shown in Eq. 3 by subtracting the average of OFF regions from the data. This calculation makes the average of OFF regions for $|\Delta S_{21}|$ centered on zero for better comparisons with ON regions.

$$|S_{nm}| = 20 \times \log_{10}^{|S_{nm}|} \quad n, m = 1, 2 \qquad (\text{Eq. 2})$$

$$|\Delta S_{21}| = |\Delta S_{21}^{raw}| - \overline{|\Delta S_{21}^{raw,OFF}|} \qquad (\text{Eq. 3})$$

## 2.6 Statistics

The minimum number of samples required for statistical comparisons was determined via power analysis with a significance level of 0.05 and a power of 0.8. Sample sizes used in this work were equal to or larger than the minimum number of samples determined by power analysis. Data were normally distributed, and therefore parametric statistics were used. Paired sample t-test was used to compare measurements taken from the same samples. Two-sample Student's t-test was used where measurements were taken from different samples.

# 3. Results

## 3.1 Optogenetic stimulation of neuronal activity.

Optical stimulation of neuronal activity via optogenetics can reliably evoke action potentials with a frequency of 10 Hz or more, and spike latencies of around 10 msec (Boyden et al., 2005). Fluorescent signal produced by a calcium indicator such as jRGECO1a represents the convolution of an action potential firing rate with a single action potential calcium transient (Yaksi and Friedrich, 2006). The rise half time of jRGECO1a is less than 0.1 sec (Dana et al., 2016). Effectively, an increase in the jRGECO1a signal can be expected to faithfully represent the initiation of an evoked action potential train with ~ 0.1 sec precision. While jRGECO1a decay is slower (half time of up to 2 sec), the start of an exponential drop-off in fluorescence signal intensity also effectively represents the cessation of an action potential train. Fluorescence of jRGECO1a was therefore used to develop an effective optical stimulation protocol and to verify the presence of evoked or spontaneous neural activity in this work. Different protocols of optogenetic stimulation of neuronal spheroids were tested to determine the optimum protocol for evoking prolonged episodes of neuronal activity. Figure S3 shows two samples of different stimulation protocols, the first one with 20 Hz pulses and a 20-second gap between each pulse train, and the second one a train of 120 pulses (Fig. 2E) for 60 seconds, repeating 4 times per experiment. The second method of stimulation led to a significant increase in neuronal activity that was sustained for the duration of the stimulation period (a representative activity trace of one culture is shown in Fig. 2E, the total number of cultures used was $n = 12$). Activity slowly recovered to the rest state after stimulation was terminated. Experiments were performed in the presence of bicuculline, a $GABA_A$ transmitter antagonist, to reduce the influence of inhibitory neurons. Experiments without bicuculline were also performed to confirm that results held in the absence of disinhibition.

## 3.2 Frequency domain analysis of transmission S-parameter changes during neuronal stimulation.

A representative sample of the measured transmission $|S_{21}|$ vs. frequency is shown in Figure S4 and a representative sample of baseline detection for $|S_{21}^{raw}|$ and calculation of $|\Delta S_{21}^{raw}|$ is shown in Figure S5. Figure 3A shows a heatmap of $|\Delta S_{21}^{raw}|$ (colormap shows the magnitude) per frequency (vertical axis) versus time (horizontal axis). The same time axis was used to plot the $\Delta F/F$ trace of the same culture, with stimulation periods indicated by blue lines on top of the $\Delta F/F$ plot. ON and OFF regions, used for comparing periods of stimulation and rest, respectively, are marked as shaded rectangular regions (purple and green). Each stimulation period was considered as an event. Figure 3B shows the summary of frequency domain analysis of ON and OFF $|\Delta S_{21}|$ data (n = 48 events). The thick line and shaded area show the average +/-standard deviation. A significant increase in transmission ($|\Delta S_{21}|$) was observed for ON relative to OFF data. Comparison of ON and OFF data in 2 GHz bins (last bin is 1 GHz between 8 to 9 GHz), as well as results of t-test for each bin ($n = 48$ events), are shown in Figure 3C. Significant increases in transmission for ON vs. OFF data were found in all frequency bins. The differences between transmission S parameters in quiescent (rest) and stimulated periods are the first evidence of the dependency of electrical properties of neurons in the gigahertz frequency range on activity.

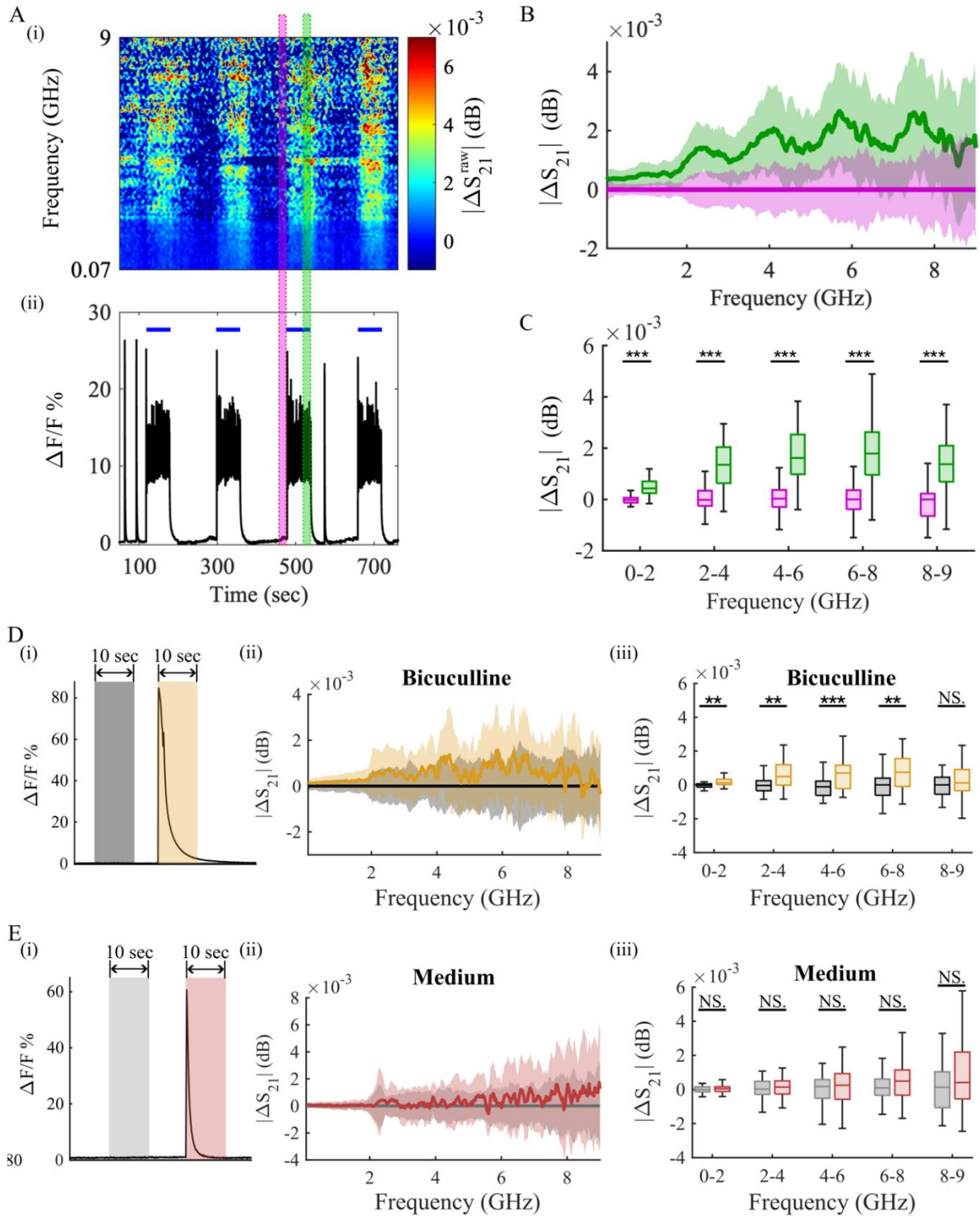

**Figure 3**. Frequency domain analysis of transmission coefficient for ON/OFF regions. **A**. (i) A heatmap of $|\Delta S_{21}|$. (ii) The $\Delta F/F$ trace of neural activity is plotted on the same time axis as the heatmap in (i) to compare the optical and electrical data together during rest and stimulation. Stimulation periods are indicated by a blue line above the $\Delta F/F$ trace. **B**. Frequency spectrum analysis of $|\Delta S_{21}|$ measured during ON (green) and OFF (purple) time periods. **C**. Data from panel B was binned as indicated, *** represents $p < 0.001$ (Student's t-test). **D**. (i) A representative $\Delta F/F$ trace with analysis windows containing periods of quiescence (grey) and spontaneous bursting (orange), in the presence of bicuculline. (ii) Corresponding frequency domain analysis of $|\Delta S_{21}|$, and (iii) comparison of data in frequency bins, using Student's t-test, *** $p < 0.001$, ** $p < 0.01$, NS. - not significant. **E**. Analysis of periods of quiescence (grey) and spontaneous bursting (light red) in regular culture medium, NS. - not significant.

Neurons in cultured spheroids spontaneously and synchronously fire bursts of action potentials, similar to activity states in the developing brain (Ming et al., 2020). These bursts cause brief, paroxysmal increases in [$Ca^{2+}$], and in turn, jRGECO1a fluorescence (Hasan et al., 2019). Spheroids on test chips also fired these spontaneous bursts, both in regular culture medium, and in medium containing bicuculline (GABA$_A$ receptor antagonist that reduces inhibition in neural networks, thereby increasing activity and synchronization, (Kemp et al., 1986)) (Fig. S2A(i)). To determine whether intracellular changes due to these spontaneous and physiological bursts of activity caused detectable changes in EM transmission, S parameters during periods of spontaneous burst activity were compared to S parameters during quiescent periods. Windows of 10 second duration were chosen for this analysis (Figure 3D(i), 3E(i), Figure S6). Significant differences were found between spontaneous bursts and quiescent periods in the presence of bicuculline (Figure 3D(ii, iii)), but not in regular culture medium (Figure 3E(ii, iii)), using $n = 35$ analysis windows. These results indicate that spontaneous bursts of neuronal activity generated detectable changes in neuronal electrical properties in the MHz-GHz frequency range. Differences were more significant in the presence of bicuculline. This was likely due to the higher amplitude of spontaneous bursting induced by this antagonist of inhibitory receptors.

## 3.3 Pharmacological suppression of neuronal activity eliminates the differences in electrical transmission.

Another comparison was conducted between ON and OFF |$\Delta S_{21}$| in cultures in regular culture medium, medium containing bicuculline, and medium containing tetrodotoxin (TTX, voltage-gated sodium channel blocker that suppresses the spontaneous and evoked action potentials (Tukker et al., 2023)). The presence of bicuculline increased the neuronal response to stimulation, while the presence of TTX suppressed the evoked response relative to that in the regular culture medium (see representative ΔF/F traces in Figure 4A). Pharmacologically-induced changes in neuronal activity were echoed by relative changes in |ΔS21|: slightly higher differences between ON and OFF regions were observed in the presence of bicuculline relative to the regular culture medium, while no differences between these regions were observed in the presence of TTX (Figure 4B). Results in Figure 4C show significant changes during stimulation (ON) compared to rest state (OFF) in the presence of bicuculline ($p < 0.001$, Student's t-test, $n = 12$) and in regular culture medium ($p < 0.001$, Student's t-test, $n = 12$), but not in the presence of TTX ($p = 0.94$, Student's t-test, $n = 12$). Furthermore, |$\Delta S_{21}$| values during ON regions in bicuculline and regular culture medium were significantly higher than those in TTX (Figure 4D). These data suggest that TTX silencing of neuronal activity resulted in no electrical transmission changes during the stimulation period compared to rest. This provides further evidence that changes in |$\Delta S_{21}$| that occur during stimulation depend on the evoked neuronal activity (specifically, evoked action potentials).

## 3.4 Time domain analysis of transmission changes

Time domain analysis shows how |$\Delta S_{21}$| transmission coefficient is changing as a function of time, starting with the rest state, during stimulation, and then recovering back to rest level. Based on frequency domain analysis, |$\Delta S_{21}$| for frequencies from 2 to 8 GHz was averaged per time point, to obtain a single |$\Delta S_{21}$|(t) vector per event. Figure 5A(i) shows the average +/- standard deviation of |$\Delta S_{21}$| vs. time ($n = 48$ events in 12 cortical aggregate cultures) for an event, including the stimulation period as well as rest periods before and after stimulation. The average

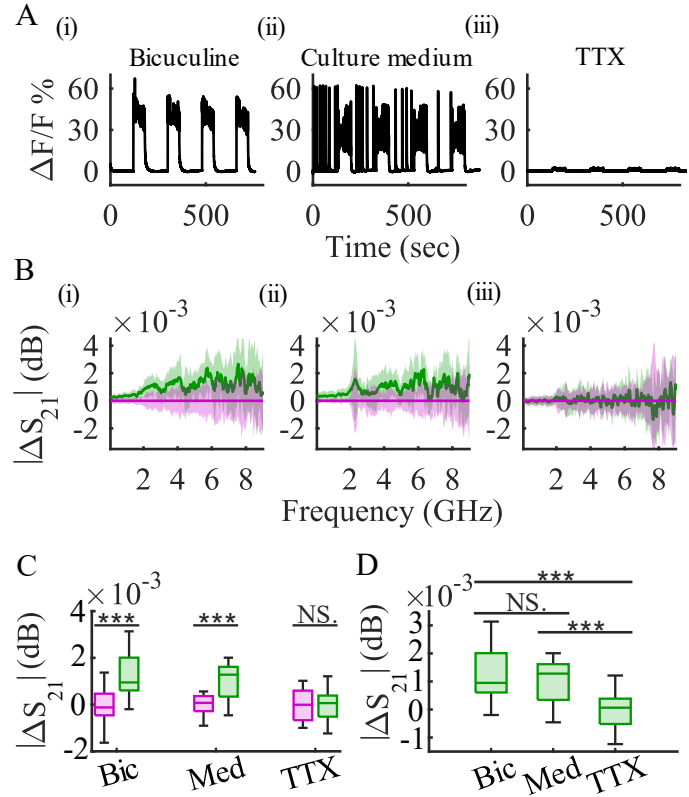

**Figure 4**. Pharmacological inhibition and changes in S21 transmission. **A**. Representative ΔF/F traces in the presence of (i) bicuculline, (ii) regular culture medium, and (iii) TTX. **B**. Analysis of |ΔS$_{21}$| ON and OFF in frequency domain, corresponding to treatment groups in (A). C. Comparison of 2-8 GHz data between ON and OFF regions in each treatment group. D. Comparison of ON region data between 3 treatment groups. *** $p < 0.001$, N.S. – Not Significant.

ΔF/F +/- standard deviation for the same events ($n = 48$ events) is plotted on the same time axis (Figure 5A(ii)) for comparison, with the start and end of the stimulation period indicated on the plot. |ΔS$_{21}$| increased gradually during stimulation to reach a maximum state (plateau). Once stimulation was completed, |ΔS$_{21}$| decreased to below resting level (undershoot) and then recovered back to baseline. This analysis showed that the change in transmission parameter: 1) is a relatively slow phenomenon, 2) follows an exponential-like trend for stimulation and post-stimulation periods. We compared |ΔS$_{21}$| baseline (OFF1, 20 seconds before start of stimulation) with maximum level reached during stimulation (ON, last 20 seconds of stimulation) and minimum reached during undershoot after stimulation ended (OFF2, 20 seconds), as indicated in Figure 5B(i). ON level was significantly higher than either OFF or OFF2 levels ($p < 0.001$, Student's t-test, $n = 48$ events), and OFF2 was significantly lower than OFF1 ($p < 0.006$, Student's t-test, $n = 48$ events), shown in Figure 5B(ii).

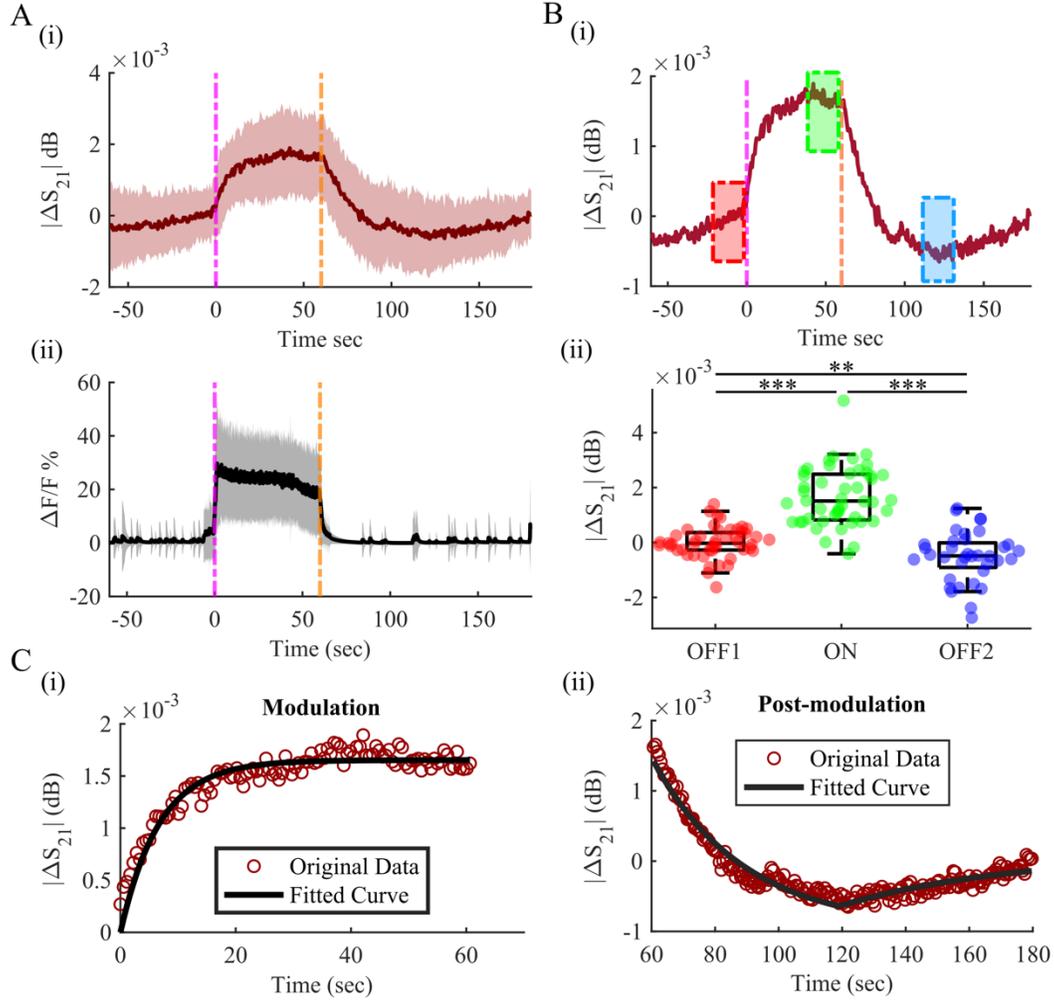

**Figure 5**. Time domain analysis. **A**. (i) $|\Delta S_{21}|$ changes during an event, starting from one minute before the start of stimulation (magenta line) (rest) and ending two minutes after the end of stimulation (orange line). The thick line represents the average, while the shaded region represents +/- standard deviation. (ii)$\Delta F/F$ trace (average is represented by the thick line and +/- standard deviation is represented by the shaded region) on the same time axis as (i). **B**. (i) Regions used for comparison of different states are indicated on the average $|\Delta S_{21}|$ trace (same as Fig. 5A(i)): red box indicates OFF1 region, representing as pre-stimulation rest state (baseline), green box shows ON stimulation region, and blue box shows OFF2 post-stimulation undershoot. (ii) Individual event data and box plots showing median +/- quartiles are plotted for regions indicated in (i). **C**. shows Actual data (red circles) and black line (fitted model) for (i) stimulation (t = 0 sec on this plot represents the start of stimulation) and (ii) post-stimulation (stimulation ended at t = 60 sec) regions. -*** $p < 0.001$, ** $p < 0.01$

## 3.5 Exponential models of $|\Delta S_{21}|(t)$

Exponential models were fitted to each stimulation and post-stimulation period, using MATLAB. Initial estimates were made for the equations and parameters in the curve fitting toolbox, and then, a nonlinear least-square solver was used to optimize the parameters. Eq. 4 and 5 show the fitted models (Table 1 shows the parameter values for Eq. 4 and 5). Figure 5C shows the actual data (red circles) and the fitted models (black lines) for each period.

$$\Delta S_{stimulation} = a \times \left(1 - e^{-\frac{t}{\tau_1}}\right) \tag{Eq. 4}$$

$$\Delta S_{post-stimulation} = b \times e^{-\frac{t}{\tau_2}} + bias_1 + \left(c \times e^{\frac{t-t_d}{\tau_3}} + bias_2\right) \times u(t - t_d) \ , u(t - t_d) = \begin{cases} 0, & t < t_d \\ 1, & t \geq t_d \end{cases} \quad (Eq. 5)$$

**Table 1.** Parameter values for exponential models of $|\Delta S_{21}|(t)$.

| Model | Parameters |
|---|---|
| $\Delta S_{stimulation}$ | a = 0.0017     $\tau_1$ = 6.765 sec |
| $\Delta S_{post-stimulation}$ | b = 0.183   c = - 0.001   bias1 = - 0.001   bias2 = 0.001<br>$\tau_2$ = 29.98 sec   $\tau_3$ = 40.00 sec   $t_d$ = 119 sec |

$\Delta S_{stimulation}$ can be accurately represented (coefficient of determination $r^2$ = 0.90, Figure 5C(i)) by a single exponential model that reaches a plateau toward the end of the stimulation period. $\Delta S_{post-stimulation}$ model consists of two exponentials, one accounting for decay after the end of stimulation and the other accounting for recovery from undershoot to the rest state baseline after a time delay ($t_d$) (coefficient of determination $r^2$ = 0.94 between model and experimental data, Figure 5C(ii)). The parameters in Table 1 are: $\tau_1$ = time constant of $\Delta S_{stimulation}$, $\tau_2$ and $\tau_3$ = time constants of $\Delta S_{post-stimulation}$ for the decaying and recovering exponential terms, $t_d$ = delay constant for the recovering exponential part, and the rest of the parameters are constants in the curve fitting.

### 3.6 Significant changes observed at resonance frequencies used for fMRI imaging

fMRI is a method that detects the blood-oxygen-level-dependent (BOLD) changes to monitor neural activity (Roth, 2023). The BOLD effect is due to the imbalance in the concentration of oxygenated hemoglobin in the activated brain regions (Greve, 2011). The main components of the fMRI system are a magnet that produces the main magnetic field (B), gradient coils, and radiofrequency (RF) coils to transmit an RF field ($B_1$) at the resonance frequency of the hydrogen atom in the main magnetic field (B) (Kwok, 2022). The RF pulse frequency used for fMRI imaging is based on the Larmor equation for the atom of hydrogen, and it could be as low as 64 MHz (at 1.5 Tesla) (~ 64 MHz) and up to 500 MHz (11.7 Tesla). To investigate neural activity-dependent changes in transmission at fMRI-relevant frequencies, we calculated $|\Delta S_{21}|$ in the 64-500 MHz range. Both $|\Delta S_{21}|$ and $\Delta F/F$ are plotted (average +/- standard deviation) during stimulation and pre and post-stimulation rest states in Figure 6A. Interestingly, analysis of $|\Delta S_{21}|$ in frequency and time domains in this frequency range shows significant changes during stimulation (Figure 6B and 6C). Figure 6B(i) shows the regions of pre-stimulation (OFF1) in red, stimulation (ON) in green, and post-stimulation (OFF2) in blue on the same time axis as Figure 6A(i), and Figure 6B(ii) shows the box plot and each data point for these regions. Significant differences were found between rest states and stimulation ($n$ = 48, Student's t-test), although no significant differences were found between the two rest states (OFF1 and OFF2). Figure 6C(i) shows the $|\Delta S_{21}|$ ON and OFF data for this frequency range, and Figure 6C(ii) shows ON and OFF data for each of the most used frequencies in fMRI studies (1.5T = 64 MHz, 3T = 128 MHz, 5T = 212 MHz, 7T = 298 MHz, 9.4T = 400 MHz, 10.5T = 447 MHz, and 11.7T = 500 MHz). These data suggest that electrical properties of neuronal tissue in the MHz frequency range are affected by neuronal activity.

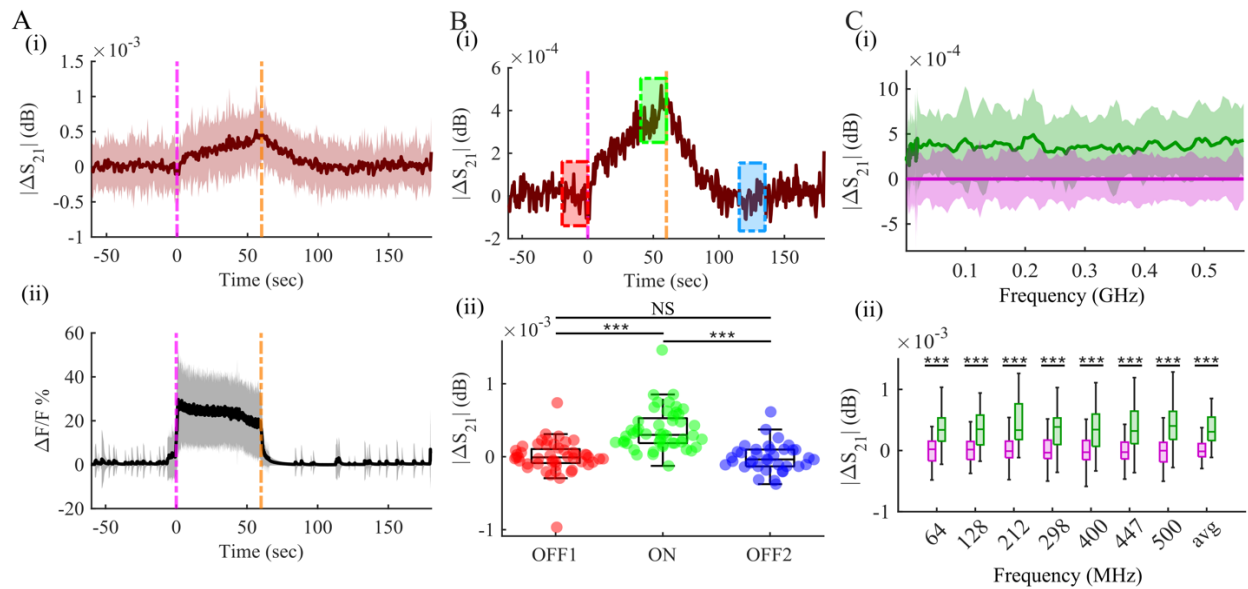

**Figure 6**. Frequency and time domain analysis in the range of MRI frequencies. **A**. for (i) |ΔS$_{21}$| and (ii) ΔF/F trace plotted on the same time axis. **B**. Regions used for comparison of different states are indicated on the average |ΔS$_{21}$| trace (same as Fig. 6A(i)): red box indicates OFF1 region, representing pre-stimulation rest state (baseline), green box shows ON stimulation region, and blue box shows OFF2 post-stimulation undershoot. (ii) Individual event data and box plots showing median +/- quartiles are plotted for regions indicated in (i), *** represents p < 0.001, NS = not significant. **C**. (i) |ΔS$_{21}$| during ON (green) and OFF (magenta) periods is plotted for frequencies between 64 and 500 MHz. (ii) ON and OFF data plotted per frequency (*** *p* < 0.001)

## 4. Discussion

Transmission parameter S$_{21}$ that we have measured in this work depends on the amount of EM power transferred from the input port through the GCPW to the output port. It therefore depends on the impedance of the GCPW, which can be modeled as a series of lumped impedances. The uninsulated portion of the GCPW (under the 'contact zone' in Fig. 1A(iv)), which is the sensing portion of the device, can be represented as a single lumped impedance, in series with lumped impedances representing the rest of the waveguide. The impedance of GCPW depends on the waveguide geometry and on the dielectric permittivity of the material above and below the waveguide (Simons, 2004). Waveguide geometry impacts the depth of penetration of the electric field of the guided EM wave into the material. Depth of penetration depends for the most part on the gap between the center strip conductor and the co-planar grounds (side conductors). In the device reported in this work, the gap in the 'contact zone' is 1 µm, which means that most of the electric field (and therefore power) is contained within 1-2 µm of the GCPW plane. The dielectric permittivity of the silica below the GCPW is constant, while any changes in the dielectric permittivity of the sample material on top of the GCPW will affect the lumped impedance of the waveguide under the 'contact zone'. Changes in this component of impedance will in turn impact wave propagation by changing the proportion of wave power that was transmitted or reflected. These changes are frequency-dependent, as can be readily seen in Fig. 1(E) when comparing transmission and reflection between dry and medium-filled conditions in the same device. The cortical spheroid placed on top of the GCPW active zone flattens (without losing its 3D layering) over time, such that during the time of the measurement, a depth of 1-2 µm above the waveguide surface contains both flattened neuronal soma and high-density of axons and dendrites (neurites, corresponding to neuropil of the intact cortex), as shown in Fig. 2 (A, B). Activity-dependent changes in the dielectric permittivity of this neuronal material are the cause of the activity-dependent changes seen in the transmission parameter ($|\Delta S_{21}|$).

Reliable detection of activity-dependent $|\Delta S_{21}|$ depended on robust, repeatable activation of neuronal activity. Cortical neurons in the relatively small spheroids used in this work form an extensive network of reciprocal synaptic connections with each other. This network is relatively easily stimulated into population activity, which is the synchronous firing of action potentials by most neurons in the network. This enabled us to stimulate neuronal material above opaque GCPWs by delivering optical pulses to the exposed portion of the network (Fig. 2). Recurrent synaptic connectivity ensured that the entire network in the plated spheroid was activated. We reasoned that changes in the dielectric properties of neural tissue that govern its interaction with MHz and GHz electric fields will occur at relatively slow time scales, since synaptic vesicle depletion and replenishment, accumulation and clearance of intracellular $Cl^-$, pH changes, and water transport all have time constants on the scale of seconds (Raimondo et al., 2015; Staley et al., 1998). We therefore developed a stimulation protocol that caused a sustained increase in neural activity with a duration of up to 1 minute in plated cortical spheroids. Repeated neural events of this duration enabled us to measure $S_{21}$ changes which occurred at slow time scales. Remarkably, detectable changes in $S_{21}$ were also caused by spontaneous, synchronized burst activity. This suggests that intracellular changes caused by these bursts, which are similar to activity patterns in the developing cortex (Ming et al., 2020), last for several seconds after cessation of the burst. Promisingly, these results also suggest that the sensing method developed in this work may be able to measure intracellular changes caused not just by prolonged artificial stimulation, but also by physiological activity.

Biological mechanisms responsible for detected changes in dielectric properties must: (1) affect interactions between cortical neurons and electric fields at MHz – GHz frequencies, and (2) occur at relatively slow time scales with time constants in the 6 – 40 second range found in this work. Activity-dependent neuron swelling and reduction in extracellular space volume may be one such mechanism. While free water dispersion occurs at 25 GHz at 37 °C, bound water (water that is hydrating proteins or in the immediate vicinity of various cellular membranes) has dispersion reaching into the low GHz range. Movement of free water in and out of neurons may impact bound water interactions with cellular material, and cause activity-dependent changes in this frequency range. Changes in cellular and extracellular volume due to neural activity occur on a time scale of seconds (Holthoff and Witte, 1996; Syková et al., 2003), similar to the time scale of the detected signal in this work. Another mechanism may be the endocytosis-dependent replenishment of synaptic vesicles in presynaptic terminals after activity-dependent exocytosis and neurotransmitter release. This process occurs with a time constant of 15 seconds (Granseth et al., 2006), and may be accompanied by significant, and slowly decaying (with time constants of 10 – 20 sec) elevations of presynaptic $Na^+$ after episodes of high neural activity (Zhu et al., 2020). Cortical synaptic vesicles have a radius of approximately 20 nm. Electrical polarization due to membranes (Maxwell-Wagner effect) varies with frequency depending on the radius of membrane-enclosed particles: smaller particles interact with electric fields at higher frequencies (Foster and Schwan, 1989). Brain tissue, containing synaptic vesicles, was predicted to have the Maxwell-Wagner effect contributing to polarizability at frequencies up to 1 GHz (Foster et al., 1979). Progressive depletion and slow replenishment of synaptic vesicles in cortical spheroids during and after evoked activity periods may explain the significant changes we observed at frequencies below 1 GHz. It is not clear what mechanism is responsible for the below-baseline dip in the $S_{21}$ parameter after activity and subsequent recovery (Fig. 5), although endocytosis overshoot at pre-synaptic terminals has been described (Xue et al., 2012). Finally, slow changes in proteins and ionic concentrations cannot be ruled out as potential mechanisms.

Sensing method developed in this work may be effective in human neural tissue as well as rat neural spheroids. While the human brain is substantially larger than the rat brain and possesses a different macro-architecture, there are strong similarities at the micro-scale that characterize the GCPW effective sensing area (a few microns above the surface of the waveguide). The intracellular mechanisms described above are present in human as well as rat neurons and could be expected to cause similar changes in tissue dielectric properties.

Established label-free methods could be broadly grouped into: (1) methods that directly sense neural activity, such as conventional extracellular and intracellular electrophysiology, electroencephalography, and magnetoencephalography, and (2) methods that sense consequences of neural activity, such as fMRI, fNIRS, and intrinsic signal imaging (ISI). The only direct method that is capable of sensing intracellular changes is intracellular electrophysiology. However, this method is invasive at the cellular level, and causes perturbation of the cell's contents. Indirect methods are based on hemodynamics, and reflect changes in blood due to neural activity (Ferrari and Quaresima, 2012; Formisano and Goebel, 2003; Grinvald et al., 1986). Our method is capable of sensing activity-dependent changes in spheroids which lack vascularization, and are composed predominantly of neuronal-lineage cells (Hasan and Berdichevsky, 2021). Therefore, the detected signal has a neuronal origin. Since the cell membrane is effectively 'invisible' to EM fields in the frequency range used in this work, the detected signal does not originate from membrane channels. This differentiates the sensing approach used in this work from conventional electrophysiology that relies on membrane ion channels. Spheroids are characterized by a lack of significant extracellular spaces (Hasan and Berdichevsky, 2021), similar to intact gray matter. This suggests that most of the detected signal is due to intracellular changes, although we cannot rule out water transport from extracellular to intracellular spaces. The time scale of the detected signal is also substantially different between this method and conventional electrophysiology. Electrical activity detected by the latter is characterized by rapid onset and termination (on the scale of tens of milliseconds in neural spheroids (Hasan et al., 2019)), reflecting membrane dynamics. In contrast, time constants of changes in EM transmission were on the order of seconds to tens of seconds. This reflects a different nature of the biological processes that are producing these changes.

Significant neural activity-dependent changes were detected at MHz frequencies that correspond to radio frequency (RF) pulses used in magnetic resonance imaging (MRI). These effects at Larmor frequencies for typical magnetic field strengths used in MRI are plotted in Figure 6. As can be seen in this figure, as well as Figures 3B and 3C, the effect size is smaller than what we measured for GHz frequencies but still significant. In MRI, the useable signal is generated through the interaction of externally applied RF pulses with proton magnetization (Buxton, 2002). Our findings suggest that interaction between external RF fields and neural tissue depends on activity-dependent changes in tissue dielectric properties. This previously unexplored mode of interaction may have significance for further development of functional MRI.

## 4. Conclusion

The key finding of this work is that neuronal activity causes changes in the high-frequency electrical properties of neuronal tissue. These changes may be caused by several putative intracellular mechanisms linked to several brain disorders.  Label-free sensing utilized in the work may be a promising approach for gaining a better understanding of the mechanisms of brain disorders, and discovery of novel treatments.  The sensing device used in this work placed neuronal tissue into close contact with GCPWs transmitting MHz-GHz frequency EM waves.  Activity-dependent changes in tissue properties caused changes in transmission of the EM waves, enabling detection.  Neural spheroids were used as a model of neural tissue. Facile control of neuronal activity in spheroids by optogenetic and pharmacological means enabled validation of detected changes in transmission.  Simultaneous detection of neuronal activity and high-frequency transmission enabled investigation of the dynamics of changes in transmission, and revealed the presence of relatively slow, activity-dependent intracellular processes with time constants in the seconds - tens of seconds range.  Neural spheroids are finding increased applications in drug development as models of neurological and psychiatric disorders (Strong et al., 2023).  The device developed in this work may therefore find application in investigation of the effects of novel drugs and other therapeutic agents on intracellular mechanisms linked with brain disorders.

**CRediT authorship contribution statement**

**Saeed Omidi**: Investigation, Methodology, Formal analysis, Visualization, Writing – original draft, Writing – review and editing.  **Gianluca Fabi**: Investigation, Formal analysis, Writing – original draft. **Xiaopeng Wang**: Investigation, Methodology.  **James C. M. Hwang**: Conceptualization, Funding acquisition, Methodology, Project administration, Supervision. **Yevgeny Berdichevsky**: Conceptualization, Funding acquisition, Methodology, Project administration, Supervision, Writing – original draft, Writing – review and editing.

**Declaration of competing interest**

Authors have no competing interests to declare.

**Acknowledgments**

This work was supported by the Air Force Office of Scientific Research (AFOSR) FA9550-19-1-0419 and National Science Foundation (NSF) NCS ECCS 1835278.

**Appendix A. Supplementary Data**

Supplementary data are in a separate document.

**Data availability**

Data will be made available on request

**References**

Auer, T., Schreppel, P., Erker, T., Schwarzer, C., 2020. Impaired chloride homeostasis in epilepsy: Molecular basis, impact on treatment, and current treatment approaches. Pharmacol. Ther. 205, 107422. https://doi.org/10.1016/j.pharmthera.2019.107422

Barbati, S.A., Podda, M.V., Grassi, C., 2022. Tuning brain networks: The emerging role of transcranial direct current stimulation on structural plasticity. Front. Cell. Neurosci. 16, 945777. https://doi.org/10.3389/fncel.2022.945777

Bounik, R., Cardes, F., Ulusan, H., Modena, M.M., Hierlemann, A., 2022. Impedance Imaging of Cells and Tissues: Design and Applications. BME Front. 2022, 9857485. https://doi.org/10.34133/2022/9857485

Boyden, E.S., Zhang, F., Bamberg, E., Nagel, G., Deisseroth, K., 2005. Millisecond-timescale, genetically targeted optical control of neural activity. Nat. Neurosci. 8, 1263–1268. https://doi.org/10.1038/nn1525

Brewer, G.J., Torricelli, J.R., Evege, E.K., Price, P.J., 1993. Optimized survival of hippocampal neurons in B27-supplemented Neurobasal, a new serum-free medium combination. J. Neurosci. Res. 35, 567–576. https://doi.org/10.1002/jnr.490350513

Buxton, R.B., 2002. Introduction to Functional Magnetic Resonance Imaging: Principles and Techniques. Cambridge University Press, Cambridge. https://doi.org/10.1017/CBO9780511549854

Calabresi, P., Di Lazzaro, G., Marino, G., Campanelli, F., Ghiglieri, V., 2023. Advances in understanding the function of alpha-synuclein: implications for Parkinson's disease. Brain 146, 3587–3597. https://doi.org/10.1093/brain/awad150

Dana, H., Mohar, B., Sun, Y., Narayan, S., Gordus, A., Hasseman, J.P., Tsegaye, G., Holt, G.T., Hu, A., Walpita, D., Patel, R., Macklin, J.J., Bargmann, C.I., Ahrens, M.B., Schreiter, E.R., Jayaraman, V., Looger, L.L., Svoboda, K., Kim, D.S., 2016. Sensitive red protein calcium indicators for imaging neural activity. eLife 5, e12727. https://doi.org/10.7554/eLife.12727

Dingle, Y.-T.L., Boutin, M.E., Chirila, A.M., Livi, L.L., Labriola, N.R., Jakubek, L.M., Morgan, J.R., Darling, E.M., Kauer, J.A., Hoffman-Kim, D., 2015. Three-Dimensional Neural Spheroid Culture: An In Vitro Model for Cortical Studies. Tissue Eng. Part C Methods 21, 1274–1283. https://doi.org/10.1089/ten.tec.2015.0135

Du, X., Ferguson, C., Ma, X., Cheng, X., Hwang, J.C.M., 2022. Ultra-Wideband Impedance Spectroscopy of the Nucleus in a Live Cell. IEEE J. Electromagn. RF Microw. Med. Biol. 6, 267–272. https://doi.org/10.1109/JERM.2021.3121258

Eilers, P.H., Boelens, H.F., 2005. Baseline correction with asymmetric least squares smoothing.

Elghajiji, A., Wang, X., Weston, S.D., Zeck, G., Hengerer, B., Tosh, D., Rocha, P.R.F., 2021. Electrochemical Impedance Spectroscopy as a Tool for Monitoring Cell Differentiation from Floor Plate Progenitors to Midbrain Neurons in Real Time. Adv. Biol. 5, 2100330. https://doi.org/10.1002/adbi.202100330

Ermolina, I., Hayashi, Y., Raicu, V., Feldman, Y., 2015. Proteins in Solutions and Natural Membranes, in: Raicu, Valerica, Feldman, Y. (Eds.), Dielectric Relaxation in Biological Systems: Physical Principles, Methods, and Applications. Oxford University Press, p. 0. https://doi.org/10.1093/acprof:oso/9780199686513.003.0011

Farokhniaee, A., McIntyre, C.C., 2019. Theoretical Principles of Deep Brain Stimulation Induced Synaptic Suppression. Brain Stimulat. 12, 1402–1409. https://doi.org/10.1016/j.brs.2019.07.005

Ferguson, C., Pini, N., Du, X., Farina, M., Hwang, J.M.C., Pietrangelo, T., Cheng, X., 2021. Broadband electrical impedance as a novel characterization of oxidative stress in single L6 skeletal muscle cells. Anal. Chim. Acta 1173, 338678. https://doi.org/10.1016/j.aca.2021.338678

Ferguson, C.A., Hwang, J.C.M., Zhang, Y., Cheng, X., 2023. Single-Cell Classification Based on Population Nucleus Size Combining Microwave Impedance Spectroscopy and Machine Learning. Sensors 23, 1001. https://doi.org/10.3390/s23021001

Ferrari, M., Quaresima, V., 2012. A brief review on the history of human functional near-infrared spectroscopy (fNIRS) development and fields of application. NeuroImage 63, 921–935. https://doi.org/10.1016/j.neuroimage.2012.03.049

Formisano, E., Goebel, R., 2003. Tracking cognitive processes with functional MRI mental chronometry. Curr. Opin. Neurobiol. 13, 174–181. https://doi.org/10.1016/s0959-4388(03)00044-8

Foster, K.R., Schepps, J.L., Stoy, R.D., Schwan, H.P., 1979. Dielectric properties of brain tissue between 0.01 and 10 GHz. Phys. Med. Biol. 24, 1177. https://doi.org/10.1088/0031-9155/24/6/008

Foster, K.R., Schwan, H.P., 1989. Dielectric properties of tissues and biological materials: a critical review. Crit. Rev. Biomed. Eng. 17, 25–104.

Granseth, B., Odermatt, B., Royle, S.J., Lagnado, L., 2006. Clathrin-mediated endocytosis is the dominant mechanism of vesicle retrieval at hippocampal synapses. Neuron 51, 773–786. https://doi.org/10.1016/j.neuron.2006.08.029

Greve, J.M., 2011. The BOLD effect. Methods Mol. Biol. Clifton NJ 771, 153–169. https://doi.org/10.1007/978-1-61779-219-9_8

Grieco, S.F., Holmes, T.C., Xu, X., 2023. Probing neural circuit mechanisms in Alzheimer's disease using novel technologies. Mol. Psychiatry 28, 4407–4420. https://doi.org/10.1038/s41380-023-02018-x

Grinvald, A., Lieke, E., Frostig, R.D., Gilbert, C.D., Wiesel, T.N., 1986. Functional architecture of cortex revealed by optical imaging of intrinsic signals. Nature 324, 361–364. https://doi.org/10.1038/324361a0

Hasan, M.F., Berdichevsky, Y., 2021. Neuron and astrocyte aggregation and sorting in three-dimensional neuronal constructs. Commun. Biol. 4, 1–16. https://doi.org/10.1038/s42003-021-02104-2

Hasan, M.F., Ghiasvand, S., Wang, H., Miwa, J.M., Berdichevsky, Y., 2019. Neural layer self-assembly in geometrically confined rat and human 3D cultures. Biofabrication 11, 045011. https://doi.org/10.1088/1758-5090/ab2d3f

Holthoff, K., Witte, O.W., 1996. Intrinsic optical signals in rat neocortical slices measured with near-infrared dark-field microscopy reveal changes in extracellular space. J. Neurosci. Off. J. Soc. Neurosci. 16, 2740–2749. https://doi.org/10.1523/JNEUROSCI.16-08-02740.1996

Hwang, J.C.M., 2021. Label-Free Noninvasive Cell Characterization: A Methodology Using Broadband Impedance Spectroscopy. IEEE Microw. Mag. 22, 78–87. https://doi.org/10.1109/MMM.2021.3056834

Irimajiri, A., Hanai, T., Inouye, A., 1975. Dielectric properties of synaptosomes isolated from rat brain cortex. Biophys. Struct. Mech. 1, 273–283. https://doi.org/10.1007/BF00537641

Kemp, J.A., Marshall, G.R., Woodruff, G.N., 1986. Quantitative evaluation of the potencies of GABA-receptor agonists and antagonists using the rat hippocampal slice preparation. Br. J. Pharmacol. 87, 677–684. https://doi.org/10.1111/j.1476-5381.1986.tb14585.x

Krukiewicz, K., 2020. Electrochemical impedance spectroscopy as a versatile tool for the characterization of neural tissue: A mini review. Electrochem. Commun. 116, 106742. https://doi.org/10.1016/j.elecom.2020.106742

Kwok, W.E., 2022. Basic Principles of and Practical Guide to Clinical MRI Radiofrequency Coils. RadioGraphics 42, 898–918. https://doi.org/10.1148/rg.210110

Lei, K.F., Ho, Y.-C., Huang, Chia-Hao, Huang, Chun-Hao, Pai, P.C., 2021. Characterization of stem cell-like property in cancer cells based on single-cell impedance measurement in a microfluidic platform. Talanta 229, 122259. https://doi.org/10.1016/j.talanta.2021.122259

Levy, E., Puzenko, A., Kaatze, U., Ishai, P.B., Feldman, Y., 2012. Dielectric spectra broadening as the signature of dipole-matrix interaction. I. Water in nonionic solutions. J. Chem. Phys. 136, 114502. https://doi.org/10.1063/1.3687914

Mertens, M., Chavoshi, M., Peytral-Rieu, O., Grenier, K., Schreurs, D., 2023. Dielectric Spectroscopy: Revealing the True Colors of Biological Matter. IEEE Microw. Mag. 24, 49–62. https://doi.org/10.1109/MMM.2022.3233510

Ming, Y., Hasan, M.F., Tatic-Lucic, S., Berdichevsky, Y., 2020. Micro Three-Dimensional Neuronal Cultures Generate Developing Cortex-Like Activity Patterns. Front. Neurosci. 14.

Raimondo, J.V., Burman, R.J., Katz, A.A., Akerman, C.J., 2015. Ion dynamics during seizures. Front. Cell. Neurosci. 9, 419. https://doi.org/10.3389/fncel.2015.00419

Roth, B.J., 2023. Can MRI Be Used as a Sensor to Record Neural Activity? Sensors 23, 1337. https://doi.org/10.3390/s23031337

Simons, R.N., 2004. Coplanar waveguide circuits, components, and systems. John Wiley & Sons.

Staley, K.J., Longacher, M., Bains, J.S., Yee, A., 1998. Presynaptic modulation of CA3 network activity. Nat. Neurosci. 1, 201–209. https://doi.org/10.1038/651

Strong, C.E., Zhang, J., Carrasco, M., Kundu, S., Boutin, M., Vishwasrao, H.D., Liu, J., Medina, A., Chen, Y.-C., Wilson, K., Lee, E.M., Ferrer, M., 2023. Functional brain region-specific neural spheroids for modeling neurological diseases and therapeutics screening. Commun. Biol. 6, 1–18. https://doi.org/10.1038/s42003-023-05582-8

Syková, E., Vargová, L., Kubinová, S., Jendelová, P., Chvátal, A., 2003. The relationship between changes in intrinsic optical signals and cell swelling in rat spinal cord slices. NeuroImage 18, 214–230. https://doi.org/10.1016/s1053-8119(02)00014-9

Tamra, A., Zedek, A., Rols, M.-P., Dubuc, D., Grenier, K., 2022. Single Cell Microwave Biosensor for Monitoring Cellular Response to Electrochemotherapy. IEEE Trans. Biomed. Eng. 69, 3407–3414. https://doi.org/10.1109/TBME.2022.3170267

Tukker, A.M., Vrolijk, M.F., van Kleef, R.G.D.M., Sijm, D.T.H.M., Westerink, R.H.S., 2023. Mixture effects of tetrodotoxin (TTX) and drugs targeting voltage-gated sodium channels on spontaneous neuronal activity in vitro. Toxicol. Lett. 373, 53–61. https://doi.org/10.1016/j.toxlet.2022.11.005

Vasu, S.O., Kaphzan, H., 2022. Calcium channels control tDCS-induced spontaneous vesicle release from axon terminals. Brain Stimulat. 15, 270–282. https://doi.org/10.1016/j.brs.2022.01.005

Wu, Q., Shaikh, M.A., Meymand, E.S., Zhang, B., Luk, K.C., Trojanowski, J.Q., Lee, V.M.-Y., 2020. Neuronal activity modulates alpha-synuclein aggregation and spreading in organotypic brain slice cultures and in vivo. Acta Neuropathol. (Berl.) 140, 831–849. https://doi.org/10.1007/s00401-020-02227-6

Xue, L., McNeil, B.D., Wu, X.-S., Luo, F., He, L., Wu, L.-G., 2012. A Membrane Pool Retrieved via Endocytosis Overshoot at Nerve Terminals: A Study of Its Retrieval

Mechanism and Role. J. Neurosci. 32, 3398–3404. https://doi.org/10.1523/JNEUROSCI.5943-11.2012

Yaksi, E., Friedrich, R.W., 2006. Reconstruction of firing rate changes across neuronal populations by temporally deconvolved Ca2+ imaging. Nat. Methods 3, 377–383. https://doi.org/10.1038/nmeth874

Zhu, Y., Li, D., Huang, H., 2020. Activity and Cytosolic Na+ Regulate Synaptic Vesicle Endocytosis. J. Neurosci. Off. J. Soc. Neurosci. 40, 6112–6120. https://doi.org/10.1523/JNEUROSCI.0119-20.2020

**Supplementary Methods**

**Spontaneous Activity Analysis**

Spontaneous activity in neural spheroids plated on test chips and on regular culture dishes was optically recorded as described in section 2.3 and converted into ΔF/F trace as described in section 2.4. Periods of 700 seconds were analyzed in each recording ($n = 7$ recordings on chips, and on culture dishes). The power of the neuronal activity was determined by integrating ΔF/F trace. Bursts of spontaneous activity were defined as events with significantly higher ΔF/F amplitude than background neuronal activity. The frequency of occurrence of these bursts was determined by detecting and quantifying paroxysmal ΔF/F peaks per unit time. Analysis was performed in Matlab.

**Propidium Iodide Staining and Quantification**

Propidium Iodide (PI) stains cells with compromised membrane integrity. Significant PI staining represents the presence of dead or dying cells. Neural spheroids were plated on test chips and on regular culture dishes. After 3 or 5 days post-plating (same time window as that used for measuring S parameters), PI (Biotium, 1 mg/mL in water) was diluted at 1:250 ratio with the culture medium and incubated with the cultures for 30 minutes. The fluorescent images were taken and the average intensity of PI staining per area was quantified in ImageJ ($n = 3$ spheroids for each condition).

# Supplementary Figures

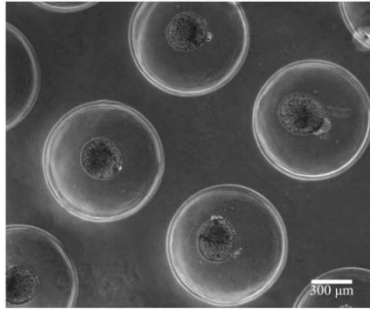

**Figure S1**. Neural spheroids formed in agarose wells.

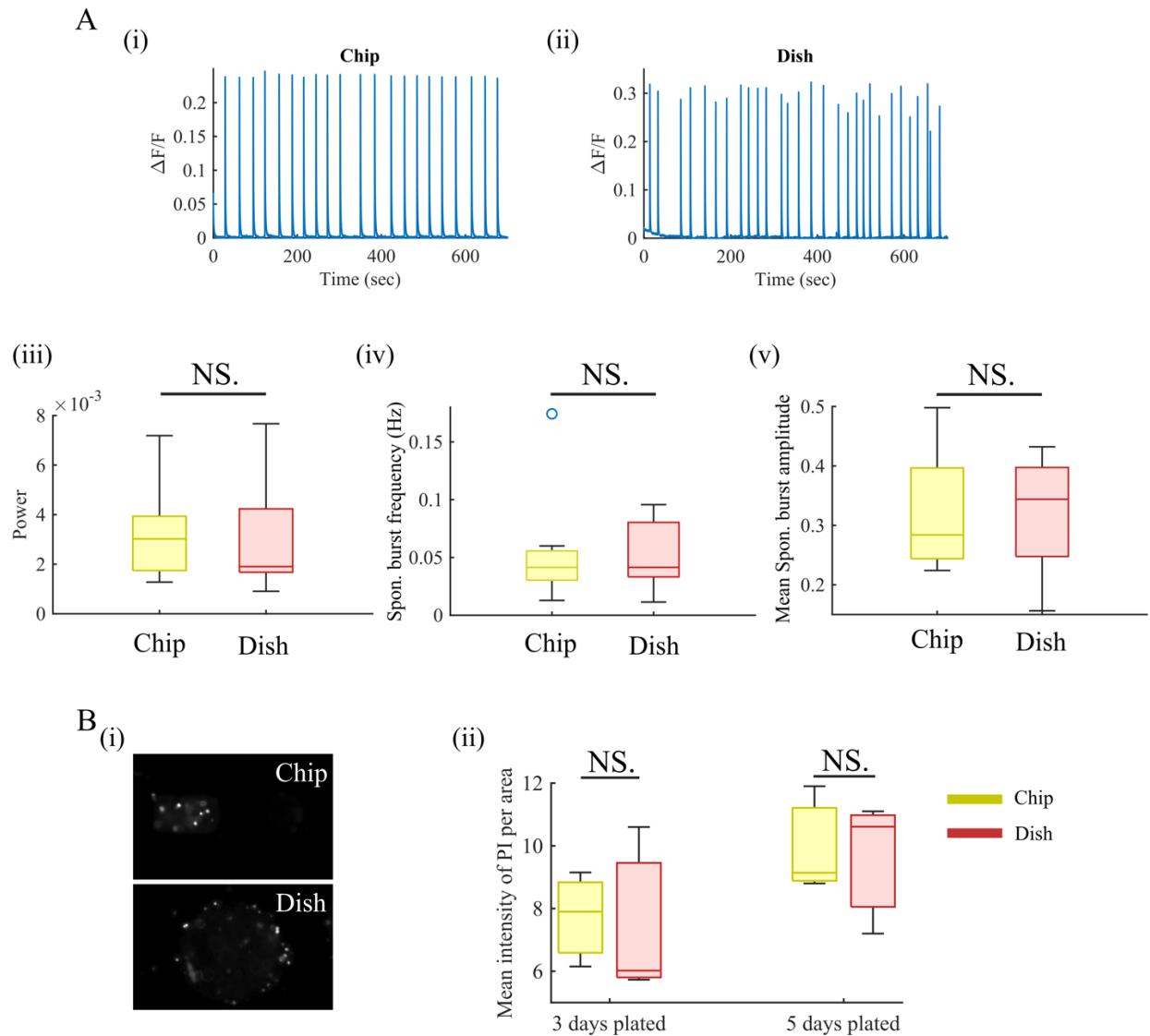

**Figure S2** Viability of neural spheroids plated on the test chips and regular culture dishes. **A**. (i), (ii) Representative ΔF/F traces showing recorded spontaneous activity in spheroids on (i) chips and (ii) culture dishes. Comparison of (iii) power of spontaneous activity, (iv) spontaneous burst frequency, and (v) spontaneous burst amplitudes. No significant differences were found between spheroids on chips and dishes. **B**. (i) representative images of PI staining, (ii) quantification of PI staining after 3 or 5 days of spheroid plating. No significant differences were found between spheroids on chips and dishes, indicating similar viability. NS. – not significant, Student's two-sample t-test.

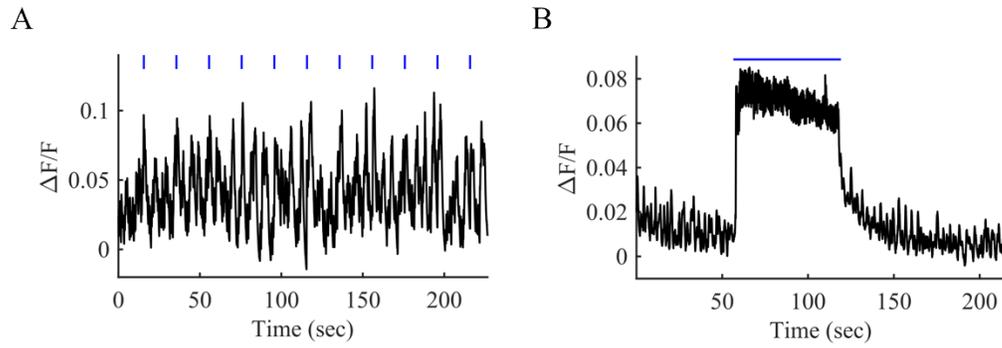

**Figure S3** Optical stimulation protocols. **A.** Panel shows a representative ΔF/F trace during stimulation with a 20 Hz pulse train that was repeated every 20 seconds. **B**. Fluorescence trace during a 2 Hz pulse train applied for 60 seconds. This stimulation protocol successfully evoked prolonged neural activity.

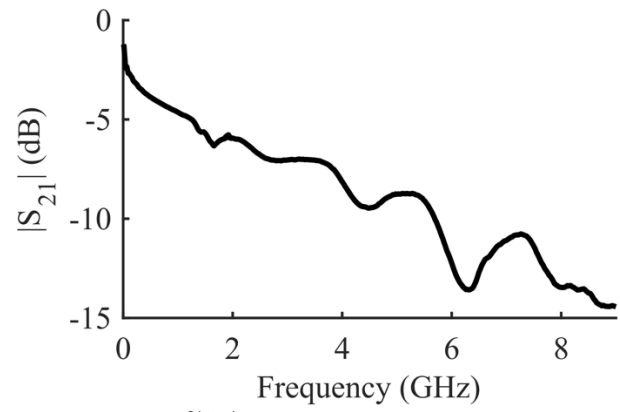

**Figure S4** Representative frequency spectrum of $|S_{21}|$.

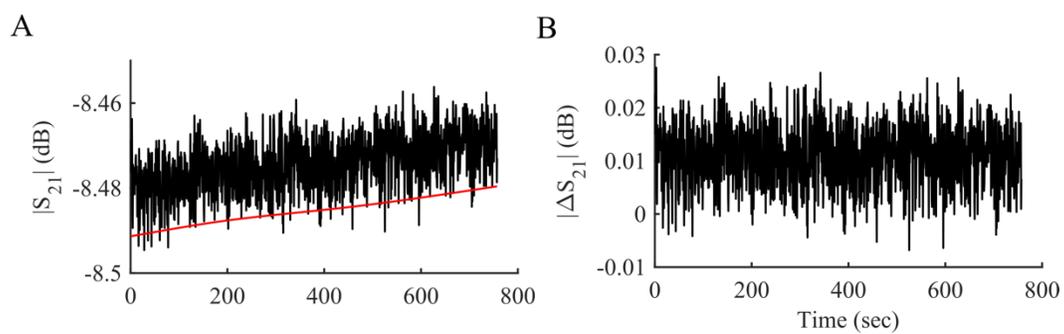

**Figure S5** Baseline detection and removal. **A**. Detection of a baseline at one frequency of $|S_{21}|$. The black trace shows the drift of $|S_{21}|$ with time, and the red trace shows the baseline determined with the asymmetric least square smoothing method. **B**. Corrected $|S_{21}|$ trace after baseline subtraction.

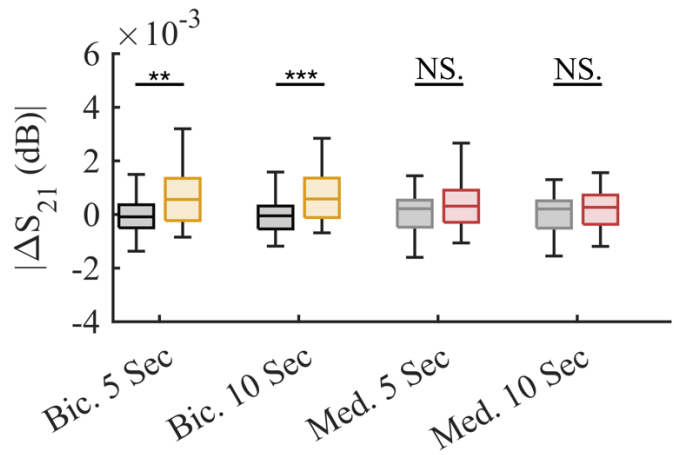

**Figure S6** Comparison of 5 and 10 second windows for analysis of changes in S-parameters during spontaneous activity. Yellow color represents data from windows with spontaneous bursts in the presence of bicuculline, red represents windows with spontaneous bursts in regular culture medium, and grey represents windows with quiescent activity. ** $p < 0.01$, *** $p < 0.001$, NS. – not significant, two-sample Student's t-test.